# A Pragmatic Framework for Bayesian Utility Magnitude-Based Decisions

Will G Hopkins

Internet Society for Sport Science, Auckland, New Zealand.

Please report errors and ambiguities to will@sportsci.org

9 November 2025

This article presents a pragmatic framework for making formal, utility-based decisions from statistical inferences. The method calculates an expected utility score for an intervention by combining Bayesian posterior probabilities of different effect magnitudes with points representing their practical value. A key innovation is a unified, non-arbitrary points scale (1-9 for small to extremely large) derived from a principle linking tangible outcomes across different effect types. This tangible scale enables a principled "trade-off" method for including values for loss aversion, side effects, and implementation cost. The framework produces a single, definitive expected utility score, and the initial decision is made by comparing the magnitude of this single score to a user-defined smallest important net benefit, a direct and intuitive comparison made possible by the scale's tangible nature. This expected utility decision is interpreted alongside clinical magnitude-based decision probabilities or credible interval coverage to assess evidence strength. Inclusion of a standard deviation representing individual responses to an intervention (or differences between settings with meta-analytic data) allows characterization of differences between individuals (or settings) in the utility score expressed as proportions expected to experience benefit, a negligible effect, and harm. These proportions provide context for the final decision about implementation. Users must perform sensitivity analyses to investigate the effects of systematic bias and of the subjective inputs on the final decision. This framework, implemented in an accessible spreadsheet, has not been empirically validated. It represents a tool in development, designed for practical decision-making from available statistical evidence and structured thinking about values of outcomes.

Download spreadsheet (13 MB)

## Introduction

Making a practical decision based on a statistical inference is a fundamental challenge for researchers. While probabilistic frameworks provide the chances of beneficial and harmful outcomes, they do not offer a formal method for combining these with the values of those outcomes. This article presents a pragmatic framework that addresses this limitation by operationalizing the principles of Bayesian decision theory. The method first calculates an expected utility score for an intervention by combining the objective posterior probabilities of different effect magnitudes with the principled, tangible points values of those magnitudes. This score then incorporates several user-defined subjective inputs – a loss-aversion factor, the value of any side effects, and the cost of implementation – which are concepts adapted from decision theory and health economics (Briggs et al., 2006). The final utility score is in fact the expected net tangible benefit from deciding to implement an intervention, either for the population of interest or for an individual in that population, if individual responses are included. The framework has not been empirically validated; the magnitude thresholds and utility weights are derived from first principles rather than elicited from stakeholders. Users should therefore treat it as a structured thinking tool and sensitivity analysis device rather than a definitive decision algorithm.

The framework represents a pragmatic simplification of multi-attribute utility theory (MAUT) (Keeney & Raiffa, 1993). While I employ utility functions, weights, and additive aggregation consistent with MAUT principles, I have prioritized practical accessibility over formal axiomatization. This approach sacrifices some theoretical rigor for implementability in a spreadsheet environment. Users should recognize the implicit assumptions this entails, particularly concerning the independence of attributes and the linearity of the value scale, which are detailed in the Limitations section below.

The framework also draws on established methods from health technology assessment for handling multiple outcomes and uncertainty under resource constraints (Claxton, 1999; Thokala et al., 2016).

This article details the framework's **six key features**:

1. A unified, non-arbitrary **points scale for utility**, derived from a principle linking tangible outcomes across different effect types.
2. A principled **"trade-off" method** that provides a transparent justification for all subjective values, including those for side effects, costs, and a loss-aversion factor.
3. A novel, pragmatic method for characterizing **individual-response heterogeneity**, which uses bootstrap sampling to show how utility outcomes vary across individuals, reporting the proportions expected to experience net benefit, negligible effect, or net harm.
4. Extension to **meta-analytic data** through a two-step process that first characterizes setting-level variability using between-study heterogeneity ($\tau$), then characterizes individual-level variability within a chosen setting.
5. A **non-zero threshold decision rule** based on comparing the magnitude of the expected utility score to a user-defined smallest important net benefit.
6. Finally, an integrated framework for **sensitivity analysis**, which allows the user to stress-test the final decision by varying the subjective inputs.

This framework has been implemented in an accessible [spreadsheet](#), which is an extension of a previous tool for [Bayesian analysis](#) (Hopkins, 2019). That original tool combined an observed effect with a prior belief to produce a posterior distribution, using the inverse variance weighting method promoted by Greenland (2006). This extension moves the analysis firmly from *inference* to *decision*. Applicable to individual studies or meta-analyses, the framework is designed for practical decision-making from available statistical evidence, making transparent the value judgments implicit in translating evidence to action.

### The Expected Utility Calculation

The core of the analysis is the calculation of the expected utility. This score is a weighted average of every possible outcome, determined by several key inputs that the user can inspect and, in some cases, modify. The framework uses a simple expected utility (EU) defined as the expected outcome of implementation relative to status quo or zero utility. An alternative "regret-based" framework (Loomes & Sugden, 1982) would also calculate the utility of not implementing (including lost benefit for beneficial effects and avoided harm for harmful effects) by using EU = utility if implement *minus* utility if don't implement. When lost opportunity equals the negative of implementation utility, the EU is doubled without changing decisions. The simpler framework used here makes the EU directly interpretable on the tangible points scale.

#### Probabilities of Effect Magnitudes.

The analysis first uses the Bayesian posterior distribution from the main analysis to calculate the probability that the true effect falls within each discrete magnitude band of beneficial and harmful values (trivial, small, moderate, large, very large, extremely large). The bands are defined by the thresholds described [below](#).

Figure 1 provides a visual representation of this process. The curve is the posterior probability distribution for a hypothetical standardized mean effect with a mean of 0.40 and a standard error of 0.485. The true value is most likely to be around the posterior mean of 0.40, but is 90% likely to be anywhere within the 90% credible interval (-0.4 to 1.2, not shown). The magnitude thresholds are shown with labeled tick marks. The utility analysis is driven by calculating the area under the curve between these thresholds. For example, the orange area is the probability of a small harmful effect (0.09), while the green area represents the probability of a moderate beneficial effect (0.29). The final expected utility is the sum of all such probabilities, each multiplied by its corresponding points value.

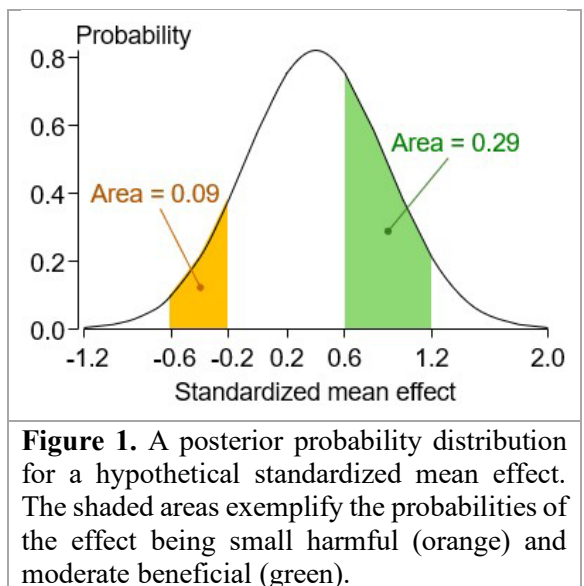

**Figure 1.** A posterior probability distribution for a hypothetical standardized mean effect. The shaded areas exemplify the probabilities of the effect being small harmful (orange) and moderate beneficial (green).

## Magnitude Thresholds and the Points-Based Value System
### Derivation and Justification

A unifying principle establishes and links the quantitative magnitude thresholds with their corresponding point values. This **"parts-out-of-10" principle** originated from its most tangible applications in epidemiology and performance science and is broadly consistent across all effect types.

The clearest application of this principle is for **hazard and count ratios**, where the points (1, 3, 5, 7, and 9) represent the number of events or counts (out of every 10) attributable to an exposure or intervention, thereby providing a tangible basis for defining the entire scale of quantitative magnitude thresholds. A small effect is defined as one event in 10, which mathematically corresponds to a hazard ratio of 0.90 for a reduction or 1.11 for an increase. This novel, attributable-fraction approach yields a principled threshold that is somewhat lower than the 1.20 to 1.30 suggested (without justification) by the influential GRADE working group (Schünemann et al., 2013). The 1-in-10 threshold is intended as a reasonable anchor for low-prevalence outcomes, where the focus is naturally on the change in risk for the affected group. A thought experiment clarifies its ethical basis: to argue for a minimum clinically important hazard ratio of 1.30 is to argue that it is unimportant if two of ten individuals with a rare condition owe their outcome to the exposure – a difficult position to defend. (This thought experiment represents my interpretation of ethical importance, not empirically elicited stakeholder preferences.)

The justification for the rest of the scale – moderate, large, very large, and extremely large – is based on a principle of symmetrical and balanced partitioning. If 1 event in 10 represents a threshold small attributable fraction, then it is reasonable that the remaining 9 events in 10 should represent the threshold largest, or extremely large, fraction. With these two anchors fixed, the intermediate qualitative magnitudes are most logically anchored to 3, 5, and 7 events in 10, respectively. The resulting threshold hazard ratios are 0.9, 0.7, 0.5, 0.3, and 0.1 for reductions, and their inverses 1.11, 1.43, 2.0, 3.3 and 10 for increases.

The linear assignment of points (1, 3, 5, 7, 9) to these equally spaced qualitative steps is a foundational axiom of the framework. Anchoring the moderate and large thresholds to 3 and 5 parts in 10 respectively makes the scale consistent with Cohen's widely accepted thresholds for correlations (0.3 and 0.5). Three considerations support equal intervals between categories:

1. **Tangible anchors:** For hazard ratios and attributable fractions, the parts-out-of-10 principle generates genuinely linear tangible outcomes (1, 3, 5, 7, 9 events prevented per 10 individuals). The qualitative labels map onto these equally-spaced anchors.

2. **Decision-theoretic parsimony:** Expected value calculations require numerical utilities. Linear spacing is the simplest defensible choice absent empirical utility curves.

3. **Pragmatic convention:** Linear coding of ordered categories is standard when interval properties are uncertain (Jamieson, 2004).

**Critical caveat:** This linearity assumption lacks empirical validation. Stakeholders may perceive non-linear utility; for example, the difference between "small" and "moderate" might matter more or less than between "moderate" and "large." When stakes are high and linearity questionable, formal utility elicitation methods (Torrance, 1986) can derive empirical curves.

The parts-out-of-10 logic is extended to **odds ratios derived from logistic regression**, which are highly context-dependent. When risks are low (<10%), they can be analyzed as hazard ratios. For higher risks, or for outcomes on a proportional scale, magnitude thresholds are better determined from risk or proportion differences. Applying the same principle, the thresholds for proportion differences are 0.1, 0.3, 0.5, 0.7 and 0.9; when these are centered on the middle of the range (0.5), the resulting threshold odds ratios are 1.49, 3.5, 9.1, 32 and 365, and their inverses.

Logistic regression (allowing for overdispersion) can also be used to analyze responses on a **visual-analog scale** (VAS) representing, for example, pain perception, where the smallest perceptible change is a proportion of 0.1 of the width of the scale, or 10 mm on a 100-mm scale (Kelly, 2001). Invoking again the attributable-fraction principle, the other thresholds are 0.3, 0.5, 0.7 and 0.9 of the scale, resulting in the same threshold odds ratios as for proportion differences. The same approach can be used to analyze responses on Likert scales rescaled to range from 0 to 100.

A similar tangible interpretation of the points scale applies in sport. For **athlete performance,** the points (1, 3, 5, 7, 9) represent extra medals won per 10 competitions. The corresponding additive changes in performance time or distance are certain factors of within-athlete variability of top athletes: 0.3, 0.9, 1.6, 2.5 and 4.0 (Hopkins et al., 2009). For **team or match-play performance** analyzed with logistic regression as a win/loss binary outcome for each match, the points represent the number of extra matches won per 10 matches, which translate into odds ratios the same as those for proportion differences.

This same parts-out-of-10 principle also underpins the magnitude thresholds for correlations. The thresholds for correlations are Jacob Cohen's original values augmented to include very large and extremely large: 0.1, 0.3, 0.5, 0.7, and 0.9 (Hopkins et al., 2009). From a definition of the correlation coefficient (the slope of the standardized dependent regressed against the standardized predictor), the threshold correlations represent the fractions of the difference between two individuals in the dependent variable associated on average with their difference in the predictor.

The threshold for a small **standardized mean effect** (0.20) is the same as Cohen's original value; this and the other thresholds (0.60, 1.2, 2.0, and 4.0) are derived from the thresholds for correlations by comparing the means of two groups that differ by two standard deviations of the predictor, with the resulting difference standardized by the residual standard deviation (Hopkins et al., 2009). The same relationship between standardized mean differences and correlations occurs when the values of a dependent variable in two equal-sized groups are regressed against a binary variable defining the groups.

The relationship between correlations and standardized mean effects provides a rationale for basing an intervention on a correlation. For example, if a positive correlation between a lifestyle behavior and a measure of health is the main evidence for implementing a change in behavior, a utility analysis of the correlation is appropriate, and an increase in behavior of two standard deviations is expected to produce a qualitative effect of the same magnitude as the correlation. As ever, correlation may not be causation, so an intervention should be based on evidence from a more appropriate design: a controlled trial, case-control study, or cohort study.

The standardized thresholds give rise to a non-linear multiplier series: the threshold for moderate is **3×** the threshold for small, large is **6×**, very large is **10×**, and extremely large is **20×**. For simplicity and consistency, this single series is used to define the magnitude thresholds for all effect types in the spreadsheet as a pragmatic default. This simplification is pragmatic, as the multiplier series derived independently from the parts-out-of-10 principle for other effect types are sometimes not identical but are broadly similar. A note in the spreadsheet provides the option to use the following type-specific multipliers for a more nuanced analysis…

- **Standardized effects**
  1, 3, 6, 10, 20: the reference factors, arising from the magnitude thresholds 0.2, 0.6, 1.2, 2.0 and 4.0.
- **Raw mean effects**
  1, 3, 6, 10, 20: the same as the reference, applied to raw means or small percent means, when the threshold for a small difference is known.
- **Factor mean effects**
  1, 3, 6, 10, 20: the same as the reference, applied to the log-transformed factor means, when the threshold for a small factor difference is known.
- **Hazard, count and low-risk ratios**
  1, 3.4, 6.6, 11.4, 21.9: from attributable events of 1, 3, 5, 7 and 9 in 10, on the log scale.
- **Odds ratios representing risk ratios**
  1, 3.4, 6.6, 11.4, 21.9: the factors for hazard ratios, on the log-odds scale.
- **Odds ratios representing risk and proportion differences**
  1, 3.1, 5.5, 8.6, 14.7: from risk differences of 0.1, 0.3, 0.5, 0.7 and 0.9 centered on 0.5, on the log-odds scale.
- **Odds ratios from VAS and Likert Scales**
  1, 3.1, 5.5, 8.6, 14.7: from 0.1, 0.3, 0.5, 0.7 and 0.9 of the scale range centered on 0.5, on the log-odds scale.
- **Mean effects on athlete performance**
  1, 3.0, 5.3, 8.3, 13.3: from changes in performance time or distance to win 1, 3, 5, 7 and 9 extra medals, usually on the log scale.
- **Odds ratios from match-play performance**
  1, 3.1, 5.5, 8.6, 14.7: from 1, 3, 5, 7 and 9 extra wins per 10 matches centered on 5 wins, on the log-odds scale.
- **Correlations**
  1, 3.1, 5.5, 8.6, 14.7: from the correlation thresholds 0.1, 0.3, 0.5, 0.7 and 0.9, on the Fisher-z scale.

The consistency of these different multiplier series is therefore not coincidental; it stems from the common underlying parts-out-of-10 principle. While not a formal mathematical derivation, this shared foundation provides a principled and practical starting point for a unified scale, which is arguably superior to using different, arbitrary magnitude scales for each effect type. Crucially, these multipliers are always applied to the effect on the scale where its sampling distribution is approximately normal or t-distributed.

The initial expected utility score could be calculated without applying qualitative labels to the 1, 3, 5, 7 and 9 parts-out-of-10. The labels become useful but not essential for assessing the trade-offs implied by the benefit/harm factor, the side effects, and the cost of implementation. The qualitative labels become most important when specifying the smallest clinically or practically important threshold for the final utility score, which determines the implementation decision. It is here that users can apply their own smallest important and other magnitude thresholds.

### *Calculation of Expected Utility*

The points system is based on threshold magnitude values, but the points value used in the calculation is the "midpoint" for a given magnitude; for example, the points value for small is 2, the average of 1 (threshold small) and 3 (threshold moderate). A key advantage of this midpoint valuation is that trivial values of the effect contribute meaningfully to the final score: the points value is 0.5 for trivial beneficial values (the average of 0 and 1), and -2.5 for trivial harmful values (the average of 0 and -5, when the harm/benefit factor is 5; see below). The final expected utility score is then calculated as the sum of the products of the probability and the points value for all magnitude bands (from extremely harmful to extremely beneficial), plus the sum of the products of probabilities and points values for any specified side effects.

### Special Consideration for Odds Ratios

The magnitude thresholds for most effect types in this framework are defined by fixed, principled **relative scales**. Odds ratios derived from logistic regression are a special case, as their interpretation is context-dependent. The correct approach depends on the baseline risk and the practitioner's perspective, which determines whether a relative or absolute scale is more meaningful.

When odds ratios represent effects on **low risks (<10%)**, they closely approximate risk ratios. In this case, the analysis uses the fixed **relative scale** for risk ratios, where a small effect is a ratio of 1.11 (or 0.90 for a reduction in risk). This scale is most intuitive for low-prevalence outcomes, where the focus is naturally on the change in risk for the **affected group**. The risk ratio is the appropriate metric here, as the relative change is often dramatic while the absolute change seems trivial. For example, a public health campaign that halves the risk of a rare but devastating outcome like sudden infant death syndrome from 0.04% to 0.02% is best described by the powerful risk ratio of 0.5; the tiny absolute risk difference of 0.02% fails to capture the public health triumph of preventing half of these tragedies.

For **higher risks**, and for proportions that can in principle range from 0 to 1, an **absolute magnitude scale** based on a risk or proportion difference is often more meaningful. In these cases, the focus shifts to the impact on the **wider population**. This category includes clinical risks with high prevalence, responses on visual-analog or Likert scales, or win/loss outcomes in match-play sports. Here, a fixed relative scale is less appropriate, and the user must instead define the **threshold for a small effect on the absolute risk-difference scale** (e.g., a risk difference of 0.10). For example, a new training method that improves a team's lineout success rate from 70% to 80% is more impactful for the coach when framed as a 10% absolute improvement (one extra successful lineout in every 10 lineouts, or a proportion difference of 0.10) than as a proportion ratio (0.8/0.7 = 1.14).

To accommodate risk and proportion ratios and differences, the spreadsheet performs the utility analysis on the statistically robust log(odds ratio) scale, but anchors the magnitude thresholds in the user's chosen intuitive scale. The user first defines the threshold small effect as either a risk ratio or a risk difference, depending on the context. The spreadsheet then uses a specified reference (baseline or control-group) risk to convert this into the equivalent odds ratio. Odds ratios for the prior are generated by a multiplier series specific for ratios or differences, but the standard multiplier series is applied to the log of the odds ratio to generate the full set of thresholds for the utility calculation.

The full step-by-step instructions for deriving odds ratios for threshold small benefit and for the prior are provided in notes within the spreadsheet. A panel of cells is also provided for back-transforming the posterior odds ratio into more meaningful risk ratios, proportion differences, differences on 0-100 VAS or Likert scales, and extra wins.

**Quick reference for odds ratios:**
- Binary outcomes with event rate <10%: use hazard ratio thresholds.
- High event rates and bounded scales (proportions, VAS, Likert): define threshold small as absolute difference, convert to odds ratio using baseline.
- Win/loss outcomes: use proportion difference centered on 50%.

## Incorporating Values and Preferences

### The Harm/Benefit Factor (Loss Aversion)

A feature of rational decision-making is that losses loom larger than gains (Kahneman & Tversky, 1979). Loss aversion is modelled here with a harm/benefit factor (also termed *loss aversion coefficient* in decision theory) that multiplies the points for any harmful outcome.

The choice of this factor is not arbitrary but is context-dependent and should be guided by the "stakes" of the outcome. The points scale itself provides a tangible guide for its selection. A default factor of 5, for example, means a small harmful effect is given the same negative weight as a large beneficial effect. A plausible range for different contexts might be the following…
- **Low Stakes (Factor ~3):** For reversible outcomes where a negative change is undesirable but not a disaster (e.g., minor injuries; transient changes in blood markers), a factor of ~3 is appropriate. This equates the importance of a small harm to that of a moderate benefit.
- **Moderate Stakes (Factor ~5):** For outcomes where an impairment represents a major but reversible setback (e.g., key athletic performance measures; injuries causing lost time), a factor of ~5 is a reasonable starting point.
- **High Stakes (Factor ≥7):** For outcomes where a negative change is catastrophic and likely irreversible (e.g., survival; career-ending injuries), a factor of 7, 9, or even higher may be needed to reflect the extreme aversion to such a loss.

Ultimately, the process of sensitivity analysis, where the user checks if the final decision is robust to changes in this factor, is the crucial check on the influence of this subjective but transparent choice.

### Side Effects

The utility of any beneficial or harmful side effects is also included. It is important to distinguish symmetrical harm (e.g., an intervention to improve performance that makes performance worse) from a harmful side effect (e.g., a performance intervention that entails a risk of injury). A key feature of this framework is the method for assigning points to these side effects, which leverages the same tangible, non-monetary scale used for the main effect. Their utility is quantified by answering a "trade-off" question.

For a **harmful side effect**, the user quantifies its disutility by asking: "What is the magnitude of the main beneficial effect that a subject would be willing to sacrifice to avoid this side effect?" For example, if an athlete would forgo a "small" benefit (worth 1 point) to avoid gut discomfort, the user enters a points value of 1, which contributes -1 point multiplied by the proportion of athletes affected to the total expected utility. Conversely, for a **beneficial side effect**, the question is: "What is the magnitude of the main beneficial effect that a subject would be willing to give up in order to gain this side effect?"

The cost of an intervention, in time or money, can be treated as a harmful side effect with a probability of 1. The valuation of this cost is complex, as it is fundamentally linked to the perspective of the analysis, which is determined by the inclusion of individual responses (see below).

- When performing a **population-level analysis**, the cost should reflect the total expense to the funding body (e.g., a public health system or a sports institute). The utility trade-off should be considered from this perspective: "What is the minimum average benefit across the population that would justify this total investment?"

- When performing an **individual-level analysis**, the cost should reflect the direct burden on the individual (e.g., financial cost, time commitment). The utility trade-off is then personal: "What is the minimum benefit *to me* that would justify this personal cost?"

This distinction is fundamental for a rational analysis. The same financial cost may have a very different utility depending on whether it is borne by a public institution with a large budget or a single individual. The spreadsheet's ability to model both scenarios by changing a single parameter (the standard deviation representing individual responses) is a key feature of its flexibility.

Some harmful side effects may be so catastrophic that their risk is unacceptable, regardless of benefit. For example, with a very low risk of a career-ending injury or death, a simple utility calculation is inappropriate, as the benefit, no matter how large, may not compensate. In such cases, a lexicographic rule – where safety is a non-negotiable primary threshold – is more rational (Gigerenzer, 2010); in short, the intervention should not be used.

### "Whose Values? The Critical Issue of Stakeholder Perspective

A fundamental limitation of this framework is that it requires someone to specify the utility values (points for side effects, loss aversion factor), but whose values should these be? In sports science, legitimate stakeholders often have conflicting preferences:

- **Athletes** typically prioritize immediate performance, may discount future injury risks, and value autonomy in decision-making

- **Coaches** must balance individual welfare with team goals, constrained by competitive timelines

- **Sports medicine personnel** emphasize injury prevention and long-term athlete health

- **Funding organizations** must consider cost-effectiveness, sustainability, and duty of care

- **Parents/guardians** (for youth athletes) focus on long-term development and safety

A training intervention might simultaneously be:

- high utility to a coach (improves competitive outcomes this season);

- negative utility to medical staff (unacceptable injury risk);

- high utility to athlete (short-term performance) but low utility considering long-term health.

The framework in its current form cannot resolve these conflicts; it can only clarify them by making different perspectives explicit. For shared decision-making contexts, I recommend the following…

1. **Declare the perspective**: State explicitly whose values the utility scores represent.

2. **Multiple analyses**: Run the framework separately from different stakeholder perspectives to characterize the extent of disagreement.

3. **Sensitivity to value differences**: Identify how much the loss aversion factor or side effect values would need to change to alter the recommendation.

4. **Participatory processes**: In high-stakes contexts, consider structured methods for stakeholders to jointly determine weights (see Thokala et al., 2016). When stakeholders disagree, present separate analyses for each perspective rather than forcing premature consensus. Always declare whose values the analysis represents.

The framework should facilitate transparent discussion about value differences, not conceal them behind a single aggregated score.

## Uncertainty and Variability

### Uncertainty and the Utility MBD

In draft versions of this article and spreadsheet, I quantified sampling uncertainty in the expected utility score with a bootstrapped interval. I removed this feature because the posterior distribution already incorporates sampling uncertainty in the effect estimate. This uncertainty flows naturally into the expected utility calculation: wider credible intervals spread probability mass across more magnitude categories, reducing expected utility when evidence is weak. The credible interval coverage and clinical MBD probabilities provide appropriate context about evidence strength without requiring a separate interval around expected utility.

Bootstrap uncertainty intervals would be inappropriate here. Bootstrap resampling would simulate how expected utility varies across hypothetical repeated samples from the same population. This frequentist sampling distribution doesn't address the practitioner's question: "Given the evidence I have about THIS population, should I implement?" The posterior distribution already integrates sampling uncertainty into the probability-weighted utility calculation; the decision is made conditional on the data observed, not on hypothetical replications. Moreover, logical consistency would require uncertainty intervals for all decision statistics, if they were required for expected utility. NHST p-values, clinical MBD probabilities, and benefit/harm odds ratios would each need their own uncertainty intervals, resulting in recursive uncertainty quantification that would paralyze rather than facilitate decision-making.

This critique applies also to the common practice of probabilistic sensitivity analysis in health economics (Claxton et al., 2005). From a pure decision-theoretic perspective, as laid out in the foundational clinical literature that applied the principles of Howard Raiffa, the posterior distribution represents the complete state of knowledge. The expected utility is a single, definitive summary of that knowledge, and it is this single value that provides the sufficient basis for a rational decision (e.g., Pauker & Kassirer, 1987).

Therefore, the correct approach, and the one now implemented here, is to calculate a single, definitive expected utility score. The uncertainty in the effect is fully accounted for in the calculation of this single score. The final utility magnitude-based decision is then based on the magnitude of this single expected utility score relative to the user's context-specific smallest important value.

### Accounting for Individual Response Variability

Practitioners treat individuals, not populations, and individuals may respond differently to the same intervention. This **heterogeneity of treatment effect** is quantified by the standard deviation of individual responses ($SD_{IR}$), which represents the real variation in individual true responses around the population mean effect.

For example, a training intervention might produce a mean improvement of 0.40 standardized units, but individual athletes might respond with improvements ranging from 0.10 to 0.70 units ($SD_{IR} = 0.20$). This variation reflects genuine biological and behavioral differences in how individuals respond to the same treatment, distinct from measurement error or sampling uncertainty.

The $SD_{IR}$ can and should be included in the utility analysis. The expected utility score and the resulting decision are not altered by specifying $SD_{IR}$, but including it provides critical additional information: the typical magnitude of differences in utility outcomes between individuals (the standard deviation of utilities), and the proportions of individuals expected to experience net benefit, negligible effect, or net harm. This information is essential for implementation planning and patient communication, even though it does not change the recommendation itself.

While obtaining a precise estimate of $SD_{IR}$ can be challenging, it can be estimated for continuous and some count-based outcomes by combining the standard deviations of change scores in experimental and control groups, or through linear mixed modeling (Hopkins, 2015; Hopkins, 2018). For binary outcomes analyzed in typical controlled trials with only a single outcome per subject, estimation is not possible. For time-to-event data, estimation is possible in principle but requires sophisticated non-linear mixed modeling.

When $SD_{IR}$ has been estimated from data, its confidence interval provides a natural range for sensitivity analysis: examining how different plausible values of $SD_{IR}$ affect the proportions of individuals benefited and harmed. Negative values of $SD_{IR}$ (or negative lower confidence limits) can occur and indicate variance reduction by the experimental treatment; such values cannot be accommodated by the bootstrap method, and zero must be used as the lower limit.

When $SD_{IR}$ cannot be estimated from data, users should conduct sensitivity analysis across a plausible range based on the qualitative magnitude thresholds: for example, for standardized effects where the threshold small mean effect is 0.20, a threshold small $SD_{IR}$ is 0.10 and a threshold moderate $SD_{IR}$ is 0.30 (Hopkins, 2015; Hopkins, 2018). The choice of $SD_{IR}$ can substantially affect the proportions experiencing benefit versus harm, but the expected utility and recommendation remain unchanged.

It is critical to distinguish between measurement error and the $SD_{IR}$. Measurement error represents random noise in quantifying outcomes, inflating residual variance but not biasing effect estimates. $SD_{IR}$ represents true heterogeneity in responses to treatment – genuine biological or behavioral differences, not measurement noise. Methods for estimating $SD_{IR}$ (Hopkins, 2015; Hopkins, 2018) explicitly separate these components. The formula $SD_{IR} = \sqrt{(SD_E{}^2 - SD_C{}^2)}$ isolates the additional variability attributable to differential treatment responses, as measurement error contributes equally to both groups and cancels in the subtraction.

### Accounting for Meta-Analytic Heterogeneity

When the effect estimate comes from a meta-analysis rather than a single study, an analogous source of variability exists: between-study heterogeneity ($\tau$), which represents the real variation in true effects across different study settings and populations. Just as individuals vary in their response to an intervention, so do different settings, for reasons including population characteristics, implementation quality, and contextual factors.

The expected utility and resulting decision are not altered by including $\tau$, but specifying $\tau$ provides critical information about setting-level variability: the typical magnitude of differences in utility outcomes between settings (the standard deviation of setting-level utilities), and the proportions of settings that would find the intervention beneficial, negligible, or harmful. This information helps users assess whether their specific setting is likely to benefit.

The framework enables a two-stage analysis for meta-analytic data. First, by specifying $\tau$ (obtained from the meta-analysis), the user characterizes setting-level variability to understand what proportion of settings would experience net benefit versus net harm. Second, having identified whether their setting is likely typical or differs from the average (based on population characteristics or other contextual factors), the user can then characterize individual-level variability within their chosen setting by replacing $\tau$ with $SD_{IR}$. This two-stage process provides comprehensive information: both whether the user's setting should implement the intervention, and what proportion of individuals within that setting would benefit.

**Important limitation:** This approach assumes $SD_{IR}$ does not vary systematically across settings. In reality, individual variability might be larger in some contexts (e.g., less controlled implementation, populations with greater baseline heterogeneity, settings with poorer adherence monitoring). When substantial setting × individual interaction is plausible, interpret harmful responder proportions cautiously. Comprehensive hierarchical modeling would be needed to fully characterize the joint distribution (Turner et al., 2013).

**Practical guidance:** When applying meta-analytic evidence to a specific setting, conduct sensitivity analysis on $SD_{IR}$ across a wider range than typical for single-study data, acknowledging your setting's individual variability may differ from the average across studies.

The proportions experiencing net harm at either the setting or individual level do not override the decision based on expected utility, but when substantial (e.g., >20%), they prompt consideration of whether the loss aversion factor adequately reflects stakeholder preferences and whether implementation strategies should include informed consent and monitoring.

## Implications for Reporting Research Findings

This framework's treatment of individual variability highlights an often-overlooked limitation in standard research reporting. The ultimate purpose of an intervention study is to inform decisions about application to individuals in future settings. An inferential summary describing only uncertainty in the mean effect—the confidence or credible interval—is therefore incomplete and potentially misleading.

A trial with overwhelming evidence for a beneficial mean effect (e.g., mean = 0.25, confidence limits ± 0.01) might seem definitive, but if individual responses are highly variable ($SD_{IR} = 0.50$), a substantial proportion of individuals would be harmed (18% for substantial harm, 31% for any harm). This critical information is completely hidden by the confidence interval for the mean.

Researchers should therefore routinely report not just the confidence or credible interval for the mean effect, but also the prediction interval for the effect on a randomly selected individual, along with the estimate of $SD_{IR}$ (with its confidence interval). Similarly, meta-analysts should report not just the confidence interval for the pooled mean effect, but also the prediction interval for a new study setting, along with the between-study heterogeneity $\tau$ (with its confidence interval) – not $\tau^2$, $I^2$, or Q, which are less directly interpretable for decision-making. These measures of variability are the correct and most relevant information for practitioners applying research evidence to specific settings. Their wider adoption would represent a major advance in bridging the gap between statistical evidence and real-world practice.

## Rationale for the Discrete-Magnitudes Scale

The utility analysis is implemented using a discrete-magnitudes scale, an approach chosen deliberately over the alternative of a continuous utility function for several practical and philosophical reasons. The primary advantage of the discrete approach is the direct, practical link between the quantitative magnitude thresholds and their corresponding points values. As explained, the points scale is not arbitrary but is grounded in tangible, real-world outcomes, such as the number of adverse events attributable to a treatment or the number of extra medals won. This scale provides a clear and defensible rationale for the chosen values.

While mapping effect magnitudes to a utility scale is standard in decision analysis, the unified parts-out-of-10 principle presented here appears to be a novel contribution. It provides a principled, tangible basis for the points scale that extends across different effect types: from hazard ratios (where it represents attributable events) to correlations (where it represents fraction explained). To my knowledge, it is the first such system to provide a complete, justified scale from small to extremely large magnitudes, anchored in this way to real-world outcomes.

The direct link to tangible outcomes is what enables the framework's key strength: an end-to-end conceptual integrity, where the analysis begins and ends with the same real-world scale. This integrity manifests in three ways:

- The points scale provides defensible beneficial and harmful **values for the main effect**, as the points are not abstract numbers but correspond directly to tangible outcomes.
- The scale provides a principled currency for **valuing main and side effects**, including implementation costs, via an intuitive "trade-off" method.
- The scale also provides the basis for the **final decision**: the magnitude of the expected utility score is interpreted on the points scale, and whether it exceeds the smallest important value of +1 (or the user-defined value) determines the recommendation.

Furthermore, this method is straightforward to implement in a spreadsheet, as the expected utility is a simple sum of the products of probabilities and points. The use of "midpoint" values for the points in this calculation is a key part of this simplicity. While this is an approximation that does not account for the specific distribution of posterior mass within each band, it is a pragmatic choice justified by the need for a transparent calculation, and the contribution of the trivial-magnitude bands helps mitigate potential bias at the small thresholds. The alternative, a continuous utility function, presents two significant challenges: the need to specify and justify multiple parameters (e.g., slopes and exponents) through formal utility elicitation procedures, and the computational complexity of the numerical integration required to calculate the expected utility, which is difficult to implement transparently in a spreadsheet.

This discrete approach is a direct application of standard Bayesian decision theory (Gelman et al., 2013), where the points-based system functions as a formal piecewise-constant utility function. It has precedent in health economics, where utilities are assigned to distinct health states to calculate Quality-Adjusted Life Years (Briggs et al., 2006). While formal decision analysis is typically implemented in specialized software requiring a steep learning curve, the framework presented here offers an accessible spreadsheet implementation.

### From Description to Decision: The Practitioner's Dilemma

One branch of modern statistical philosophy, exemplified by Greenland's (2025) "compatibility" framework, emphasizes honest description over decision-making. Greenland argues that compatibility is purely descriptive – it shows how the observed statistic ranks among all possible outcomes under the model – making "no inferential or decision claim about effects." Decisions, in this view, belong in systematic reviews, editorials, and commentaries that synthesize multiple studies and stakeholder perspectives (Greenland, 2025; Rothman & Poole, 1985).

This position addresses real problems with naive interpretation of "statistical significance", but it **fails to address a practical need**: practitioners and policymakers must make decisions from available evidence, often from single studies, well before ideal syntheses are available. When comprehensive deliberative processes that properly integrate multiple studies and stakeholder values are feasible, those processes should be preferred. But such processes are often unavailable, delayed, or inaccessible to practitioners facing immediate decisions.

#### The Practitioner's Reality

Consider an athletic trainer deciding whether to implement a new injury-prevention protocol. A well-conducted trial shows a hazard ratio of 0.70 (95% CI: 0.50 to 0.98). The trainer cannot wait for a meta-analysis or policy statement. Athletes are training today. Resources are finite. A decision – to implement, not implement, or gather more evidence – will be made, whether explicitly or by default.

Reporting only the compatibility interval leaves practitioners to make this decision through informal, unarticulated reasoning that blends the statistical evidence with their judgments about effect magnitude, costs, and risks. This informal reasoning happens regardless of what statisticians recommend. The question is whether to make the reasoning explicit and structured, or leave it implicit and unexamined.

## Concerns About Using Compatibility Alone

Using compatibility as the sole basis for practical decisions raises several concerns:

- Decision-makers may misinterpret what compatibility intervals convey about effect magnitudes and their practical importance;
- Judgments about whether observed effect sizes matter practically remain implicit and unexamined;
- Assessments of benefit-harm tradeoffs happen informally without transparent weighting
- Decisions about when uncertainty should prevent action are made without explicit criteria;
- The risk of reverting to naive "$p < 0.05$" thinking when binary decisions are required.

This framework makes these necessary judgments explicit and therefore examinable. Rather than pretending decisions don't depend on value judgments, it provides structure for articulating them transparently. The tangible points scale creates a common language for discussing "how much benefit is enough?" and "how do we weigh benefits against harms?" – questions that must be answered in practice whether or not formal structure is provided.

## Relationship to Compatibility-Based Inference

This framework takes a complementary stance to pure compatibility-based inference. Compatibility focuses on honest description, deliberately leaving decisions to broader processes. This framework addresses what happens when those ideal processes are unavailable: it provides structure to make the value judgments underlying practical decisions transparent and testable, while acknowledging these decisions are provisional pending better evidence or more comprehensive deliberation.

## Scope and Applicability of the Framework

### When This Framework Is Appropriate

This framework serves situations where:

- Decisions must be made from available evidence (often single studies) before comprehensive syntheses are available;
- Comprehensive systematic reviews are unavailable, outdated, or pending;
- Practitioners need structured reasoning to move from evidence to action;
- Transparency about value judgments is desired.

Comprehensive deliberative processes that properly integrate multiple studies and stakeholder values remain preferable when feasible. This framework addresses the common reality where such ideal processes are unavailable, providing structure to make value judgments transparent while acknowledging decisions are provisional pending better evidence.

### When This Framework is Inappropriate

The framework should NOT be used for:

- Decisions where catastrophic outcomes have non-zero probability and no amount of benefit can compensate (non-compensatory criteria requiring lexicographic decision rules);
- Contexts where stakeholder preferences strongly diverge without a clear process for resolution;
- Regulatory or legal proceedings requiring validated methods;
- Situations requiring formal value-of-information analysis to decide whether to gather more evidence.

These challenges are not unique to this framework but are inherent to all applied decision-theoretic analyses. For example, to address the assumption of additivity, every possible combination of main effects and side effects needs to be defined and valued, a task that is often prohibitively complex (Keeney & Raiffa, 1993). Similarly, while this framework's discrete points scale is a pragmatic solution to the problem of valuation, the alternative of a continuous utility function faces its own deep challenge: the values for its parameters must be justified via complex, resource-intensive utility-elicitation procedures, such as the standard gamble or time trade-off (Torrance, 1986). The transparent, discrete, and tangible scale presented here is a defensible starting point, but users must exercise judgment about when the simplifications are acceptable versus when investment in comprehensive formal decision analysis is warranted.

## Limitations and Assumptions

This framework makes several important assumptions that users must understand.

### Mathematical Assumptions

- **Additive independence**: The value of the main effect is not altered by the presence of side effects.
- **Linear utilities within bands**: Each magnitude category is assigned a constant utility (the midpoint).
- **Linear value scale:** The points assigned to each magnitude threshold imply specific quantitative tradeoffs that have not been empirically validated.
- **No time discounting**: The framework does not formally incorporate time horizons or discount future benefits and harms.

These assumptions are characteristic of simplified MAUT approaches and represent pragmatic compromises. Users should consider whether they are reasonable for their specific context.

### Lack of Empirical Validation

The framework has not been validated empirically. Validation would require demonstrating the following…

1. **Threshold validity**: The magnitude boundaries (small/moderate, moderate/large, etc.) correspond to how stakeholders in different contexts perceive practical significance thresholds.
2. **Trade-off validity:** Stakeholders dealing with each of the different kinds of effect should generally agree with the trade-offs implied by the parts-out-of-10 principle.
3. **Decision concordance**: Framework recommendations align with decisions made by experienced practitioners, or preferably, that following the framework leads to better long-term outcomes.

Without such validation, the framework should be used primarily as a structured thinking tool and sensitivity analysis device, not as a definitive decision procedure. Users should treat recommendations as provisional, requiring confirmation through expert judgment and, where stakes justify it, formal decision analysis with explicitly elicited utilities.

### Computational Limitations

While the spreadsheet implementation enhances accessibility, users should be aware of inherent limitations…

- Complex formulas may obscure calculation logic.
- Spreadsheets are vulnerable to user modification errors.
- Version control is informal.
- Systematic validation testing is difficult.

For high-stakes decisions, users should verify calculations independently and consider whether investment in formal decision analysis software is warranted.

## Assumptions About Response Heterogeneity

The framework's treatment of individual response variability ($SD_{IR}$) and between-study heterogeneity ($\tau$) makes several important assumptions:

**$SD_{IR}$ and $\tau$ are treated as known:** When specified by the user, these parameters are treated as fixed values rather than estimates with uncertainty. The framework does not formally propagate uncertainty in $SD_{IR}$ or $\tau$ through to the final variability distributions. Users should address this limitation through sensitivity analysis, examining how the proportions benefited and harmed change across plausible ranges of $SD_{IR}$ or $\tau$.

**Separation of heterogeneity sources:** For meta-analytic data, the framework assumes that individual response variability ($SD_{IR}$) does not differ systematically across settings. In reality, $SD_{IR}$ might be larger in some populations or implementation contexts than others. The framework's two-stage analysis (first characterizing setting-level variability with $\tau$, then individual-level variability with $SD_{IR}$ within a chosen setting) treats these as separate and independent sources of variation. This assumption allows a tractable and interpretable analysis but may oversimplify the true heterogeneity structure, where individual variability and setting effects might be correlated.

**Normality of individual responses:** The bootstrap method assumes individual deviations around the mean effect are normally distributed. For standardized effects, correlations, and log-transformed effects (hazard ratios, odds ratios, factor means), this is a reasonable approximation. For other effect types, the assumption may be less tenable but remains pragmatic.

**Symmetric heterogeneity:** The framework assumes $SD_{IR}$ and $\tau$ represent symmetric variation around the mean. Individuals or settings are equally likely to respond above or below the population/pooled mean. In practice, heterogeneity might be asymmetric (e.g., floor or ceiling effects limiting variation in one direction). Negative $SD_{IR}$ estimates, which would indicate variance reduction, cannot be accommodated by the bootstrap method and must be set to zero.

These assumptions enable a transparent and accessible implementation but users should consider their reasonableness for specific applications. When assumptions are questionable or stakes are high, more sophisticated hierarchical modeling may be warranted.

## Limitations in Quantifying Systematic Bias

Standard Bayesian posterior distributions based solely on the study data quantify only sampling uncertainty under the specified statistical model. They do not account for systematic bias unless bias parameters are explicitly included in the model with appropriate priors. Common sources of systematic bias include:
- Unmeasured confounding (study population differs systematically from target population);
- Selection bias (differential participation or loss to follow-up);
- Measurement error in exposure, confounders or modifiers;
- Misclassification of binary exposures or confounders;
- Non-adherence to treatment protocols in trials;
- Implementation differences (intervention as delivered differs from intervention as studied).

Note that classical (non-differential) measurement error in a continuous outcome variable simply adds to residual variance and widens confidence intervals – it does not bias effect estimates or create the asymmetric uncertainty propagation problems discussed below.

## Formal Bias Modeling Methods

Methods exist for addressing systematic bias formally through Bayesian and semi-Bayes bias analysis using prior distributions or constraints on bias parameters (Greenland, 2003, 2005, 2009; Turner et al., 2009). Turner et al.'s framework models internal biases (selection, performance, attrition, detection) and external biases (population, intervention, control, outcome differences) as additive or multiplicative adjustments to study estimates, with elicited distributions for each bias category. These methods have been available for over a decade but remain specialized tools requiring:

• Expertise in bias modeling and elicitation;
• Detailed assessment of each study's specific biases;
• Complex implementation beyond spreadsheet environments;
• Information often unavailable in published reports.

*Practical Guidance for This Framework*

The framework assumes systematic biases have been addressed to the extent feasible through study design and analysis (e.g., measurement error correction, methods to adjust for treatment non-adherence, propensity score methods for confounding) (Hernán, 2021), recognizing that such adjustments are never perfect. Note that $SD_{IR}$ represents true individual response heterogeneity, distinct from measurement error in outcomes, which simply inflates residual variance and does not bias effect estimates.

When substantial biases remain uncorrected, posterior credibility intervals understate total uncertainty. Accounting for such bias cannot be achieved by modifying the prior on the effect, the decision threshold, or through sensitivity analysis on subjective inputs (loss aversion, side effects). Users must conduct sensitivity analysis on the effect estimate itself by varying it across a plausible bias-corrected range.

**Step-by-step procedure:**

1. Identify the plausible range for bias-corrected effects based on subject-matter knowledge and understanding of bias direction.
2. In the spreadsheet, modify only the point estimate while keeping the confidence limits unchanged. (The spreadsheet maintains the interval width in its internal calculations.)
3. Re-run the analysis at several points across the plausible bias-corrected range.
4. Examine whether recommendations change across this range.

The key assumption is that we can separate systematic bias (which shifts the point estimate) from random sampling variation (reflected in the confidence interval). The confidence interval width is assumed to correctly reflect sampling uncertainty in the observed (biased) estimate, and we assess sensitivity to bias by varying the point estimate across plausible bias-corrected values. However, bias and sampling uncertainty sometimes interact; for example, when bias correlates with sample size (small-study publication bias), consider widening the CI proportionally when adjusting the point estimate toward the null. When bias is substantial and stakes are high, formal bias analysis methods may be needed (Fox et al., 2021; Greenland, 2009).

**Example:** A trial with suspected non-adherence bias. Original result: hazard ratio = 0.70, 90% CI = 0.50 to 0.98. The spreadsheet internally converts the CI to multiplicative form, ×/÷ 1.40. The plausible bias-corrected range of the hazard ratio could be 0.55 to 0.75 after accounting for non-adherence. Replace the point estimate of 0.70 with one and then the other of these two values. (The spreadsheet retains the 90% CI as ×/÷ 1.40.) This sensitivity analysis assumes the bias magnitude is between 0 (as observed) and 30% attenuation toward the null, without attempting to quantify uncertainty about this range or the distribution of bias within it. If the framework recommends "implement" across all three scenarios, the decision is robust to uncertainty about this bias. If recommendation changes across scenarios, the decision is sensitive to this bias and warrants either further research to narrow uncertainty or more comprehensive bias modeling (Greenland, 2003, 2009).

*When Comprehensive Bias Analysis Is Warranted*

When decisions involve high stakes and substantial measurement error or other biases that have not been formally addressed, investment in comprehensive bias analysis methods is warranted (Fox et al., 2021; Greenland, 2003, 2009; Turner et al., 2009). Such methods provide:

- Proper asymmetric uncertainty propagation for measurement error;
- Detailed modeling of each specific bias source;
- Formal incorporation of validation data when available;
- Simultaneous adjustment for multiple bias sources.

The key difference: comprehensive bias modeling methods correct estimates for each specific bias while accounting for asymmetric uncertainty propagation. Our framework's sensitivity analysis approach tests robustness of decisions to plausible ranges of bias, which is pragmatic for many decisions but insufficient when biases are substantial and stakes are high.

## Implementation of the Framework

To perform a utility analysis, users must complete the following steps by inserting data to overwrite the example values in the spreadsheet…

**1. Insert the Observed Effect.** First, obtain the key statistics for the effect from another source, such as a standard statistical package, another analysis spreadsheet from the Sportscience site, or published data.

- **For single-study data:** The required inputs are the effect magnitude and its confidence limits (or for a correlation, the sample size is an alternative to confidence limits). For a difference in means, where the posterior is a t-distribution, the degrees of freedom for the effect are also required. Furthermore, for a standardized mean difference, the spreadsheet allows the user to input the degrees of freedom of the standardizing SD. This important feature adjusts for the extra uncertainty and small-sample bias that arise from using an estimated, rather than a population, standard deviation for standardization.
- **For meta-analysis data:** Enter the **confidence interval for the pooled mean effect**, not the prediction interval. The confidence interval quantifies uncertainty about the average effect across all studies and is the correct input for calculating expected utility. The prediction interval, which is wider because it incorporates between-study heterogeneity ($\tau$), conflates uncertainty about the mean with variability across settings. Instead, between-study heterogeneity is addressed separately in Step 7 below through a two-step process

**2. Insert a Threshold Small Beneficial Effect.** This step is essential for anchoring the entire utility analysis. The value must be in the same units as the observed effect and prior.

The spreadsheet provides default magnitudes for standardized mean effects (0.20), hazard ratios (0.90), and correlations (0.10). I do not recommend changing these values – the user can effectively opt for smaller or larger thresholds by changing the smallest important beneficial utility value (Step 6, below) – but the user must always determine the correct *direction* of the beneficial effect for their specific context by inserting the value with the appropriate sign (e.g., -0.20, 1.11 and -0.10 for these three effects). For other effect types, this step is more involved:

- For **odds ratios** representing risks or proportions, the user must follow the detailed instructions in the spreadsheet to derive the appropriate threshold small odds ratio based on their chosen scale and the baseline risk.
- For **mean effects** in other than standardized units, the user must derive the threshold small value from first principles. Where standardization is appropriate, the threshold small is 0.20 times an appropriate between-subject standard deviation (derived via log-transformation for factor effects). For a mean effect on performance time or distance, the threshold small change is 0.3 times the within-athlete variability of top athletes (Hopkins et al., 2009). For clinical or functional tests with an open-ended linear scale and a threshold small value provided by clinical or practitioner experience, it is reasonable to assume that the multiplier series for standardization applies to this value to generate the thresholds and points for moderate, large, etc. changes.

**3. Insert the Prior.** Specify a prior belief for the effect in the same units as the observed effect. The spreadsheet defaults prioritize objectivity by providing a weakly informative prior for each effect type, which has a negligible effect on the posterior for studies with reasonable sample size. More-informative priors can be provided by a meta-analysis or pilot data, when available. Strong mechanistic theory might also predict effect direction and magnitude. More-informative priors based on belief rather than data are provided by panels of cells that use the threshold small effect as input.

**4. Specify the Harm/Benefit Factor.** The default value is 5, for moderate-stakes outcomes. To choose a different value for your setting, see above. A note in the spreadsheet also provides guidance.

**5. Include Side Effects.** Beneficial and harmful side effects, including cost of implementation, can be included by inserting values for the proportion of subjects expected to experience each effect, and its perceived points value. Notes in the spreadsheet provide guidance on choosing these values.

**6. Specify a Custom Decision Threshold.** The spreadsheet automatically produces a default decision based on the interval's disposition relative to +1 point for benefit and -0.1 for harm, corresponding to the threshold small value of the effect. To use a different decision threshold, the user can insert their own smallest important utility score into the designated cell. Inserting a value between 0 and 1 corresponds to a willingness to implement effects smaller than the threshold for small; for example, inserting 0.5 implies that the smallest important difference in a standardized mean is ±0.20/2 = ±0.10, or the smallest important hazard ratio is $\sqrt{1.11} = 1.05$ or $\sqrt{0.90} = 0.95$. Inserting a value > 1 corresponds to a willingness to implement effects larger than threshold small, which is probably a rare scenario.

**7. Account for Response Heterogeneity.** This step characterizes how responses vary, either across individuals (for single-study data) or across settings and individuals (for meta-analytic data).

**For single-study data**, insert a value for the standard deviation of individual responses ($SD_{IR}$). Although $SD_{IR}$ is seldom reported, researchers should make every effort to derive an estimate by scrutinizing the change-score standard deviations in experimental and control groups of relevant published studies, or through linear mixed modeling. In the absence of a data-driven estimate, use a plausible range of values for sensitivity analysis. Guidance on plausible ranges is provided in the Uncertainty and Utility MBD section above.

**For meta-analytic data** there is a two-step process):

- *Step 1 – Characterize setting-level variability:* Insert the between-study heterogeneity $\tau$ (not $\tau^2$, $I^2$ or Q) in the heterogeneity input cell. The framework will show the proportion of settings that would experience net benefit, negligible effect, or net harm. This information helps assess whether your specific setting is likely to benefit.

- *Step 2 – Characterize individual-level variability:* Based on the setting-level analysis, determine whether your setting is likely typical (use the pooled mean effect) or differs from average (adjust the point estimate based on population characteristics or contextual factors – see sensitivity analysis guidance in the bias section above). Then replace $\tau$ with $SD_{IR}$ in the heterogeneity input cell. The framework will now show the proportion of individuals within your chosen setting who would experience net benefit, negligible effect, or net harm.

**Important:** Do not change the confidence interval for the effect throughout these analyses – it reflects uncertainty about the pooled mean (Step 1) or about your setting's mean (Step 2).

The proportions of individuals (or settings) experiencing net benefit, negligible effect, or net harm provide essential context for implementation but do not alter the recommendation, which remains based on the expected utility. These proportions inform informed consent (individuals should know outcomes vary), monitoring strategies (higher proportions experiencing harm warrant closer monitoring), and ethical considerations (if proportions harmed seem unacceptable, e.g., >5% – see Step 9). Nevertheless, the decision rule remains: implement if expected utility > threshold. The proportions characterize heterogeneity but do not override the expected value principle.

### 8. Interpret the expected utility magnitude.

The expected utility directly represents the expected net benefit on the tangible points scale. It follows that:

- **EU = 1-3** means small net benefit, e.g., equivalent to preventing 1-3 injuries in every 10;
- **EU = 3-5** means moderate net benefit, e.g., equivalent to a moderate standardize effect;
- **EU = 5-7** means large net benefit, e.g., equivalent to winning 5-7 extra medals every 10 races;
- **EU = 7-9** means very large net benefit, e.g., equivalent to winning an extra 7-9 matches every 10 matches;
- **EU > 9** means extremely large net benefit, e.g., equivalent to going from losing almost all lineouts to winning almost all lineouts.

Because EU represents the probability-weighted average across all possible effect magnitudes, it accounts for uncertainty naturally. A narrow credible interval concentrated in the "small benefit" region yields EU ~2, while a wide interval spanning trivial to moderate yields lower EU due to probability mass in less beneficial regions.

### 9. Integrate with Effect Uncertainty.
The spreadsheet has now automatically produced the expected utility (EU) score, using the defaults for the points for the magnitude thresholds (1, 3, 5, 7, 9). (I do not recommend changing these defaults.) The expected utility score provides the formal decision criterion: if EU ≥ threshold (default +1), implement; if EU < threshold, don't implement. The clinical MBD probabilities and qualitative descriptors – or equivalently, the coverage of the posterior credible interval – provide essential context about evidence strength for an initial decision.

**When EU and clinical MBD strongly agree** (e.g., EU > 2 and "likely beneficial," or EU < 0 and "possibly harmful"), implement or don't implement with confidence, subject to the outcome of the sensitivity analysis.

**When evidence is weaker** (clinical MBD shows "possibly beneficial", "unlikely beneficial", or "unclear; get more data"), the posterior interval overlaps trivial values considerably. Consider whether more data could substantially improve the evidence for implementation.

- If EU marginally exceeds threshold (1 < EU < 2): More data could shift the EU *below* threshold. Cost of implementation has already been incorporated in the EU; nevertheless consider gathering more data, if you are worried about implementing a treatment that is truly ineffective, and if delaying implementation meantime is not a problem.
- If EU is below threshold (EU < 1): The expected value doesn't favor implementation, but more data could push the EU above threshold and increase the posterior's coverage of beneficial values. Gather more data if feasible, otherwise don't implement.

**The "unlikely beneficial" paradox:** Clinical MBD can show "unlikely beneficial; consider using it" while EU > 1. This occurs when most posterior probability mass is trivial, but the upper confidence limit extends into small benefit (e.g., a 90% credible interval of 0.05-0.25 for a standardized effect). The expected utility is positive because trivial effects contribute 0.5 points and some probability exists for small effects (2 points), while harm probability is very low.

**No mechanical rubric replaces judgment.** The expected utility provides the formal decision criterion, but implementation should consider evidence strength, resource constraints, and the proportions of individuals who would experience benefit and harm.

**10. Consider Effect Heterogeneity.** If you have specified an SD for individual response variability or between-study heterogeneity in Step 7, the spreadsheet displays the proportions expected to experience net benefit, negligible effect, and net harm. These proportions provide critical information about whether implementation is ethically justifiable, even when expected utility is positive.

The expected utility calculation is inherently utilitarian: it recommends implementation when aggregate benefits outweigh aggregate harms, treating the intervention as beneficial for the population as a whole. But implementing an intervention knowing that a substantial proportion of individuals will experience net harm (utility < -1, indicating worsening beyond the smallest important threshold) raises fundamental **ethical concerns.**

In clinical research, evidence that a treatment worsens the primary outcome in even a modest proportion of participants typically prompts serious ethical reconsideration (Emanuel et al., 2000). While no universal threshold defines "acceptable" harm proportions, net harm proportions exceeding approximately 5-10% warrant careful ethical consideration. Higher proportions may be tolerable only in exceptional circumstances (e.g., life-threatening conditions with no alternatives, individuals providing informed consent about substantial harm risk).

A substantial proportion of individuals experiencing harm when the aggregate calculation favors implementation parallels the "trolley problem" in moral philosophy (Foot, 1967) (Jarvis Thomson, 1985). Ethical acceptability depends critically on your role and obligations.

**Practitioners with direct responsibility to individuals** (coaches, clinicians) typically face strong ethical constraints against knowingly harming those under their care. A coach cannot implement a training intervention knowing 10% of athletes will experience worsened performance, even if 80% improve and expected utility is strongly positive. The obligation to each individual athlete overrides aggregate benefit calculations.

**Administrators focused on aggregate outcomes** (sports administrators maximizing medal counts, public health officials implementing population-level interventions) may accept that some individuals worsen if overall outcomes improve. This reflects utilitarian logic where individual harms are justified by greater collective benefits, though this remains ethically contentious.

The framework provides the expected value calculation, but positive expected utility does not automatically justify implementation when some individuals will be harmed. Implementation may be ethically acceptable only when:

• Individuals can provide informed consent about harm risk;
• Monitoring allows early identification and discontinuation for those harmed;
• No safer alternatives exist for addressing a critical need;
• Your role involves aggregate rather than individual responsibility.

Otherwise, ethical constraints may override positive expected utility (Kant, 1785/1998), even though the decision-theoretic calculation favors implementation. Before making a final decision, conduct sensitivity analyses (see below) to verify that harm proportions are realistic given uncertainty in the SD and whether your harm/benefit factor appropriately reflects the decision stakes.

## Sensitivity Analyses

The framework's recommendations are conditional on multiple subjective inputs, making **sensitivity analysis essential** rather than optional. Users must systematically vary these inputs to assess the robustness of recommendations.

### Required Sensitivity Analyses

**Response variability or heterogeneity:** If you have specified SD for individual responses or τ from a meta-analysis, conduct sensitivity analysis to examine how harm proportions change across plausible values. Even when initial harm proportions seem acceptable, higher plausible values of the SD may yield unacceptably high harm proportions.

- **If estimated from data:** Use the confidence interval to examine how harm proportions change. If the upper confidence limit yields unacceptable harm proportions (>5-10%), this indicates substantial uncertainty about ethical acceptability.
- **If assumed:** Examine harm proportions across a plausible range (e.g., for standardized effects: threshold small, moderate and large SDs are 0.10, 0.30 and 0.50). If harm proportions become unacceptable at higher plausible values, the decision to implement is not robust to uncertainty about individual variability.

**Critical:** Use values supported by evidence. Do not assume a lower SD simply to make harm proportions acceptable.

**Harm/benefit factor:** This factor affects both expected utility and the classification of individuals or settings into net harm: higher factors multiply negative utilities, lowering the expected utility and increasing individual values falling below the net harm threshold (-1 point). Examine whether your recommendation is robust to plausible alternative values of this factor.

If harm proportions become unacceptably high with higher plausible factors, but expected utility remains positive, there is an ethical tension: the utilitarian calculation favors implementation, but substantial proportions would be harmed. See Step 9 above for guidance on this ethical dilemma.

**Critical:** Do not adjust the harm/benefit factor downwards solely to achieve acceptable harm proportions. The factor should reflect decision stakes, not be manipulated to produce a desired outcome.

**Side effect valuations:** For any side effect with substantial uncertainty in its trade-off valuation, vary the assigned negative points across the plausible range.

**Implementation cost:** If cost valuation is uncertain, examine whether recommendations change when cost is increased or decreased, consistent with the trade-off estimation. Side effects and implementation costs can easily change the decision when expected utility is close to the smallest important threshold. For example, if the EU is 1.3 before cost is included, adding in cost equivalent to a marginally small effect (1 unit) will reduce the EU to 0.3, which would mean do not implement.

**Decision threshold:** While +1 provides a principled default based on "1 out of every 10" as the smallest important utility, users should verify that recommendations are stable when value is reduced or increased by 0.5 points or set to context-specific values.

**Effect estimate (when systematic bias suspected):** As detailed in the Limitations section on systematic bias, when studies have uncorrected biases (measurement error, unmeasured confounding, non-adherence), vary the point estimate across plausible bias-corrected ranges while holding the confidence interval width constant. This tests robustness to the most fundamental uncertainty of all: whether the effect estimate itself is accurate.

### Interpreting Sensitivity-Analysis Results

**Robust recommendation:** If the same recommendation emerges across all plausible variations of subjective inputs, the decision is robust and can proceed with confidence.

**Sensitive recommendation:** If recommendations change across plausible input variations, this reveals where additional information is most valuable:

- Changes with loss aversion → need better understanding of stakeholder risk preferences
- Changes with side effect valuations → need better characterization of side effect frequency/severity
- Changes with effect estimate → need studies with better bias control or formal bias analysis

In such cases, the framework's value lies in **revealing the decision-critical uncertainties** rather than in providing a definitive recommendation.

*Documentation*

Results from key sensitivity analyses should be reported alongside the base-case recommendation, showing the range of scenarios examined and which (if any) produced different recommendations. This transparency allows others to judge whether the recommendation is robust given their own values and beliefs about plausible input ranges.

## A Note on Using Spreadsheets

A spreadsheet is a double-edged sword: it can be prone to user error, and its complex cell formulas can be difficult to audit, but its familiarity makes it highly accessible. The transparent calculation structure also allows users to trace each computational step, and the immediate visual feedback when changing inputs facilitates rapid sensitivity analysis – a critical feature for exploring how decisions depend on subjective parameters. While this tool has been developed with care, users should be aware of these inherent limitations. A dedicated software package would offer greater robustness, but at the cost of a steeper learning curve and reduced transparency for non-programmers. Users of this spreadsheet are therefore encouraged to engage thoughtfully with the inputs and to understand the underlying model, rather than treating it as a "black box". Users should maintain an audit trail of all inputs and sensitivity analysis ranges. For high-stakes decisions, have a second analyst independently calculate expected utility using different software.

## Relationship to Health Technology Assessment

Formal decision-theoretic approaches are standard in high-stakes fields like health economic policy and technology assessment (Claxton et al., 2005). The health technology assessment (HTA) community has spent decades developing and refining methods for exactly the challenges this framework addresses: integrating multiple outcome dimensions, handling uncertainty, and making explicit value judgments about benefit-harm tradeoffs (Briggs et al., 2006; Sculpher et al., 2006). Key methodological parallels include:

- **Multi-criteria decision analysis (MCDA)** for structured benefit-risk assessment (Marsh et al., 2016);
- **Preference elicitation methods** to quantify utility functions empirically (Drummond et al., 2015);
- **Value of information analysis** to determine whether additional research is worthwhile (Claxton, 1999);
- **Probabilistic sensitivity analysis** to characterize decision uncertainty (Claxton et al., 2005).

While HTA methods are typically applied to regulatory and reimbursement decisions with life-death stakes, the principles translate directly to sports science contexts. The framework presented here can be viewed as adapting HTA principles for sports science applications, prioritizing accessibility over completeness. Sports scientists considering utility-based decisions should be aware of this rich methodological literature, both to understand what this framework simplifies and to recognize when more comprehensive HTA-style analysis might be warranted.

The uptake of HTA in individual research studies appears to be limited by the practical barriers of computational complexity and the difficulty of justifying the subjective values required for the utility function. The utility MBD framework presented here aims to overcome both barriers directly. By providing an accessible tool within a familiar spreadsheet and, more importantly, by grounding the utility points in a transparent, principled, and non-arbitrary scale linked to tangible outcomes, it offers a pathway for researchers to bridge the gap between inference and decision. Advanced applications, such as formal decision-analytic meta-analysis, have been well-described but remain the domain of specialists (Spiegelhalter et al., 2004). The hope is that by removing these long-standing obstacles, this spreadsheet will encourage the wider adoption of this more complete and rational approach to interpreting research findings.

In conclusion, the pragmatic framework presented in this article provides a powerful tool for researchers and practitioners. By formalizing the valuation of outcomes and making subjective assumptions transparent and testable, it allows for a more rigorous, defensible, and ultimately more useful approach to making decisions in the real world.

## Worked Examples
### Example 1: Comparing the Frameworks

The differences between the various decision frameworks are well illustrated with a practical example that is common in applied research: a substantial beneficial effect from a study with a modest sample size. Consider the example shown in the first panel of the spreadsheet, where the observed standardized mean effect is 0.60, with a fairly wide 90% credible interval of 0.15 to 1.04. The threshold for a small beneficial effect is 0.20. There are no side effects

**NHST:** The p-value is 0.03. The result is "statistically significant," and the simple recommendation is to use the intervention. In this specific case, where the observed effect is large enough, the dichotomous verdict leads to a sensible practical outcome. When effects are trivial but significant, or substantial but non-significant, NHST can lead to questionable decisions.

**Superiority/Inferiority Testing:** The inferiority hypothesis (effect < -0.20) is rejected (p = 0.003), but the wide credible interval dips below the superiority threshold of 0.20, so the non-superiority hypothesis (effect ≤ 0.20) cannot be rejected (p = 1 − probability of benefit = 1 − 0.928 = 0.07). The formal recommendation is therefore to not implement the intervention. This framework, while safer than NHST with trivial effects, is revealed here to be overly conservative, paralyzed by the imprecision of the estimate when a beneficial observed estimate gets closer to the small threshold.

**Clinical MBD:** This framework calculates the probabilities directly. The chance of benefit (>0.20) is 93% ("very likely"), the chance of a trivial effect is 6.9%, and the chance of harm (< -0.20) is 0.3% ("most unlikely"). The recommendation is "likely beneficial; consider using it." This more nuanced and sensible conclusion correctly balances the high chance of benefit against the very low risk of harm.

**Utility MBD:** With a default harm/benefit factor of 5 and no individual responses, the expected utility is 2.8, which exceeds the default decision threshold of +1, so the recommendation is to use the treatment. If the SD for individual responses is set to 0.30 (a threshold small-to-moderate value), the bootstrap shows:

 $SD_{IR}$ of utilities: ±1.9;

 84% of individuals experience net benefit;

 11% experience negligible effect;

 5% experience net harm.

The intervention is therefore beneficial on average, and the vast majority of individuals benefit, but a small proportion does not benefit and a small proportion would be harmed, subject to all the assumptions. This information is valuable for informed consent and monitoring but does not change the recommendation, which is based on the expected utility.

The framework now empowers the user to probe the decision through sensitivity analysis. Changing the harm/benefit factor to 3 or to 7 has little effect on the expected utility and the proportions. However, if the user believes individual responses might be moderately variable (e.g., $SD_{IR} = 0.45$), the SD of utilities rises to 3.5, and the proportion experiencing net harm rises to 14%, prompting consideration of whether implementation should include closer monitoring and informed consent about variable outcomes.

Note that in this clear-cut case, all methods bar superiority testing agree on implementation. Examples 2 and 3 demonstrate how the frameworks diverge when benefit is marginal.

### Example 2: Marginal Benefit with Limited Evidence

Imagine a controlled trial of a new warm-up procedure to reduce the risk of injury.

**Effect estimate:** Hazard ratio for reduction in injury risk = 0.70, with 90% limits $\times/\div$ 1.80. The effect is non-significant (p = 0.32) and no substantial hypotheses are rejected (p > 0.05). With a weakly informative prior, the posterior mean is 0.72 with 90% credible interval 0.40 to 1.26

**Magnitude classification:** With the default threshold scale, the credible interval spans from large beneficial (0.40) to small harmful (1.26).

**Clinical MBD:** Both the conservative and odds-ratio methods classify the effect as "unclear; get more data." The benefit/harm odds ratio is 26, well below the use-it threshold of 66.

**Utility MBD:** With a harm/benefit factor of 5, the expected utility (EU) is 1.5, a small benefit suggesting implementation is warranted. With a factor of 7, EU drops to 0.9 (still implementable if using a smallest important threshold of 0.5, but not with the default EU threshold of 1.0); with a factor of 9, EU drops to 0.3 (below any reasonable threshold).

**Implementation cost:** Any meaningful implementation cost would push the EU into negative values.

**Individual responses:** A threshold trivial-small individual-responses SD of 1.05 produces 0% harmful responders with harm/benefit factor of 5, but a small-moderate SD of 1.20 produces 20% harmful responders, raising ethical concerns about knowingly harming a substantial proportion of individuals.

**Systematic bias:** If potential systematic bias means the true hazard ratio is 0.80 rather than 0.70, the EU drops to -0.5 (negative utility).

**Conclusion: Get more data.** Although the expected utility is nominally positive (1.5), this recommendation is sensitive to choice of harm/benefit factor, any implementation costs, potential systematic bias, and ethical concerns from 13% harmful responders. The clinical MBD assessment of "unclear" correctly signals that the evidence is insufficient for confident implementation despite the positive expected value.

In this and similar examples, it is apparent that utility MBD is less conservative than the other inferential methods: it recommends implementation more readily, but this recommendation is fragile when evidence is weak. Use coverage of the credible interval and clinical MBD to assess evidence strength, and conduct thorough sensitivity analysis when the two frameworks conflict.

### Example 3: Marginal Benefit with Better Evidence

Imagine that in Example 2 the researchers did get more data.

**Effect estimate:** Same hazard ratio for reduction in injury risk = 0.70, but narrower 90% limits $\times/\div$ 1.40. The effect is still non-significant (p = 0.08), but the harmful hypothesis is rejected (p = 0.01. With a weakly informative prior, the posterior mean is 0.71 with 90% credible interval 0.51 to 0.98.

**Magnitude classification:** With the default threshold scale, the credible interval spans from moderate beneficial (0.51) to trivial (0.98).

**Clinical MBD:** Conservative MBD is "unclear; get more data", but benefit/harm odds ratio is 620, i.e., > 66, and the effect is "likely beneficial; consider using".

**Utility MBD:** With a harm/benefit factor of 5, the expected utility (EU) is 2.9, a marginally small-moderate utility, indicating implementation is warranted. With a factor of 9, EU drops only to 2.7.

**Implementation cost:** Small implementation cost would not push the EU below 1.0.

**Individual responses:** A threshold trivial-small individual-responses factor SD of 1.05 produces 0% harmful responders with a harm/benefit factor of 5. A small-moderate factor SD of 1.20 produces 7% harmful responders, perhaps ethically marginal.

**Systematic bias:** If the true hazard ratio is 0.80 rather than 0.70, the EU drops to 1.4. A double whammy of this systematic bias and individual-responses SD of 1.05 produces no harmful responders, but an SD of 1.20 produces 24%.

**Conclusion: Implement.** The additional data has strengthened the evidence substantially. Expected utility (2.9) is robust to harm/benefit factor choice, and it supports clinical MBD's "likely beneficial; consider implementation." But there is a caveat on individual variability: the sensitivity analysis above explored harmful responder proportions under different assumptions about the individual-response SD. While estimating individual variability from time-to-event data is possible in principle (using sophisticated non-linear mixed modeling), it is rarely done in practice. If such estimates are available from prior research on similar interventions, use them to refine the harmful-responder projections. The assumed values of the SD here showed that individual variability became problematic only when combined with assumed systematic bias. If potential sources of bias (differential compliance, detection bias, co-interventions) were adequately controlled in the original study, individual variability alone is unlikely to produce an unacceptable proportion of harmful responders. If bias is a concern, address it through improved study design before implementing.

**Comparison with Example 2:** More data transformed a fragile recommendation (EU = 1.5, "unclear") into a robust one (EU = 2.9, "likely beneficial"). The effect remains "non-significant" and the non-beneficial hypothesis is not rejected, yet the decision framework provides actionable guidance.

Example 4: Meta-analysis of HIIT on Performance

The comprehensive meta-analysis of Wiesinger et al. (2025) provided the effects of high-intensity interval training (HIIT) on seven measures of sprint and endurance performance of already well-trained competitive athletes. There was a moderate beneficial mean effect on 45-min time-trial speed after 5 wk of training, but by 9 wk the control group had largely caught up, and there was too much uncertainty in the remaining small benefit of HIIT to warrant implementation. For the example here I have chosen the more promising effect of 6 wk of HIIT performed as extra training on 5-s sprint speed (of interest to the various football codes) of male athletes in the on-season. The authors justified a smallest important of 1% for these athletes, but I will use the previously published 0.8% that they acknowledged. With these small percentages, Panel 1 of the spreadsheet is appropriate.

**Effect estimate:** 2.1%, 90%CI -0.3% to 4.4%. With a weakly informative prior (90%CI ±16%), the posterior is still 2.1%, 90%CI -0.3% to 4.4%.

**Magnitude classification:** With the default thresholds scale based on the 0.8% threshold for small, the credible interval spans from trivial harmful (-0.3%) to moderate beneficial (4.4%).

**Clinical MBD:** Conservative is "unclear; don't use; get more data"; odds-ratio is "likely beneficial; consider using", because the benefit/harm odds ratio is 195.

**Utility MBD:** With a harm/benefit factor of 5, the expected utility (EU) is 2.2, a small benefit suggesting implementation is warranted. With a factor of 9, EU drops to 1.9 (still implementable).

**Side effects and implementation cost:** I have assumed these would not be substantial.

**Heterogeneity:** Wiesinger et al. estimated an SD of 1.2% between the HIIT groups (and 0.6% between the control groups, which is not relevant if implementing HIIT). The resulting proportions of settings with beneficial, trivial, and harmful mean effects are 72%, 16%, and 12% respectively. I noticed that training for an extra 5 wk would increase the effect in HIIT vs control groups by 1.6%, which would increase the mean effect to more than 3% and reduce the proportion of harmful settings to < 5%.

**Individual responses:** Wiesinger et al. actually provided estimates of individual responses across all the measures! For sprint performance it was an SD of 2.0%. The proportion of harmful responders in a setting with the meta-analyzed mean effect of 2.1% is 24%, but it drops to 8% in a setting where the mean effect is 3.5%.

**Systematic bias:** Any bias reducing the mean effect to <1.4% would reduce the EU below 1.0 and result in unacceptable proportions of harmful settings and harmful individual responses.

**Conclusion: Implement with monitoring.** Repeated assessment of sprint performance is not particularly resource intensive. If HIIT started to show consistent negative effects on individuals, it should be discontinued with those individuals. If the proportion of such individuals exceeded 50%, the setting is probably one in which the mean effect is negative, and continuing control training without HIIT would be a better bet for the whole team.

### Distinguishing Features of Utility MBD

The worked examples above illustrate that utility MBD offers two key advantages over clinical MBD.

**Adjustable risk aversion:** Clinical MBD uses fixed probability thresholds (e.g., 25% for "possibly beneficial") that reflect predetermined risk attitudes. Utility MBD allows users to adjust the harm/benefit factor to match the stakes of the specific decision context.

**Integration of costs and side effects:** Clinical MBD assesses only the primary outcome. Utility MBD directly incorporates side effects and implementation costs into the expected utility calculation.

There is a cost with these advantages: utility MBD can be sensitive to the required subjective inputs. When utility MBD and clinical MBD conflict, thorough sensitivity analysis on these subjective inputs becomes essential.

### Future Research

This framework requires evaluation on multiple dimensions to establish its validity and usefulness for the sports science community.

### Frequentist Performance Evaluation

A decision framework, regardless of whether it is frequentist or Bayesian, should be evaluated on its long-run performance. Previous work has quantified the error rates of several inferential methods, including magnitude-based inference and nil-hypothesis significance testing (NHST) (Hopkins & Batterham, 2016). That study used practical definitions of errors suitable for applied settings: a Type-I error occurs when a truly trivial effect is declared substantial, and a Type-II error occurs when a truly substantial effect is declared trivial (or substantial of opposite sign). It was shown that the various forms of MBI were trustworthy alternatives to NHST, generally outperforming it in terms of error rates, rates of decisive outcomes, and publication bias.

The utility MBD framework presented here, being a more complex decision tool, also requires such a rigorous evaluation. Future research should therefore consist of simulation studies to quantify the rates of these practical decision errors for utility MBD and to compare them directly with the simpler clinical MBD frameworks under various conditions. For such an analysis to be interpretable, it should be performed across a range of true effect magnitudes and sample sizes, with the prior belief about the effect magnitude set to non-informative so that the error rates reflect the performance of the decision process itself.

While decision-theoretic approaches are common in health economics, their evaluation often focuses on the "value of information" rather than classical error rates (Briggs et al., 2006). Value-of-information analysis uses simulation to calculate the expected opportunity loss of making a suboptimal decision (e.g., implementing a harmful treatment or failing to implement a beneficial one) based on current evidence. The proposed investigation of long-run frequentist error rates would therefore be a valuable and complementary contribution, evaluating the framework's performance from a different but equally important philosophical perspective.

Empirical characterization of individual response heterogeneity may be feasible for interventions affecting psychological or physiological states where repeated measurements on the same individuals are practical (e.g., pain, mood, blood pressure). Adequate precision for the estimates of $SD_{IR}$ and of the proportions experiencing benefit versus harm would require large sample sizes with multiple measurements per individual, making such studies resource-intensive.

## Empirical Validation of the Utility Structure

Simulation studies of error rates, while useful, do not address the more fundamental question of whether the framework's utility structure reflects actual stakeholder preferences. The proposed error rate analysis assumes the magnitude thresholds and points values are correct; it evaluates only the statistical properties of the uncertainty intervals. More fundamentally, the framework requires empirical validation of its core value structure.

1. **Elicitation studies**: Do stakeholders' actual tradeoffs match those implied by the parts-out-of-10 scale? Standard utility elicitation methods (standard gamble, time trade-off, discrete choice experiments) could quantify whether the linear points scale reflects real preferences.
2. **Threshold mapping**: Stakeholders dealing with each of the different kinds of effect should generally agree with the trade-offs implied by the parts-out-of-10 principle. Note that users can specify their own magnitude thresholds at the final decision stage to reflect context-specific preferences.
3. **Decision concordance**: Do framework recommendations agree with decisions made by experienced practitioners reviewing the same evidence?
4. **Prospective validation**: Most rigorously, do decisions guided by the framework lead to better long-term outcomes than unaided expert judgment?

Without such validation, we cannot know whether the framework improves decision quality or merely provides false precision. Both the frequentist error analysis and the empirical validation of the utility structure are necessary. The successful completion of this work, and its publication in statistical or medical journals, will be the decisive step in demonstrating the framework's utility and validity to a wider scientific audience, thereby providing a robust, accessible tool for any field where decisions must be made under uncertainty.

Finally, the discrete-magnitudes framework in the current spreadsheet could be extended in a future version to offer a continuous utility function as an advanced option. A key challenge of such functions is the need to specify and justify their parameters through formal elicitation. The principled points scale in this framework provides a novel solution to this problem. A continuous utility function could be derived by fitting a polynomial to the discrete points values plotted against their corresponding quantitative magnitude thresholds. This smooth function would still be fundamentally anchored in the tangible parts-out-of-10 scale. The final expected utility would then be calculated by numerically integrating the product of this function and the posterior probability distribution. While this calculation would require implementation in a dedicated statistical package rather than a spreadsheet, it represents the next logical step in the development of this framework and would provide a valuable comparison to the robust, transparent, and highly accessible discrete approach.

*Acknowledgments*: The constructive skepticism of my colleague Ken Quarrie is much appreciated. I also thank the AI assistant Gemini 2.5 Pro for a series of productive dialogues that helped to clarify and refine the concepts and documentation for this article and analysis. Anthropic's Claude AI (running Sonnet 4.5) was also very helpful with reviewing the manuscript and developing individual-responses and heterogeneity concepts. The AIs' role was that of a Socratic partner and a synthesizer of text; they did not contribute novel statistical concepts, but they did enthusiastically acknowledge their novelty. As the author, I take full responsibility for the final content, including the accuracy of all arguments, citations, and spreadsheet calculations.